\documentclass[twocolumn,tighten]{aastex6}
\usepackage{eufrak}

\newcommand{\gyr}{{\rm{Gyr}}}
\newcommand{\myr}{{\rm{Myr}}}
\newcommand{\henon}{H\'enon}
\newcommand{\cmc}{{\tt CMC}}
\newcommand{\pc}{{\rm{pc}}}
\newcommand{\msun}{{M_\odot}}
\newcommand{\bse}{{{\tt BSE}}}
\newcommand{\sse}{{{\tt SSE}}}
\newcommand{\tdelay}{{t_{\rm{delay}}}}
\newcommand{\mtot}{{M_{\rm{tot}}}}
\newcommand{\mchirp}{{M_{\rm{chirp}}}}

\newcommand{\Mpc}{{\rm{Mpc}}}
\newcommand{\yr}{{\rm{yr}}}
\newcommand{\sfr}{{\rm{SFR}}}
\newcommand{\dens}{{M_\odot \rm{Mpc}^{-3}}}
\newcommand{\kms}{{\rm{km\,s}^{-1}}}
\newcommand{\mergerrate}{{\rm{Gpc}^{-3}\rm{yr}^{-1}}}
\newcommand{\tlookback}{{t_{\rm{lb}}}}
\newcommand{\fgw}{{f_{\rm{GW}}}}
\newcommand{\trelax}{{t_{\rm{relax}}}}
\newcommand{\metal}{{\rm{Z}}}
\newcommand{\redshift}{{z}}
\newcommand{\Zmean}{{\overline{\metal}}}
\newcommand{\Zsun}{{\metal_\odot}}
\newcommand{\zform}{{\redshift_{\rm{form}}}}

\shorttitle{Dynamically Forming GW151226-like BBH Mergers}
\shortauthors{Chatterjee et~al.}
\begin{document}
%
\title{Dynamical Formation of Low-Mass Merging Black Hole Binaries like GW151226}

%
\author{Sourav Chatterjee\altaffilmark{1}, Carl L.~Rodriguez\altaffilmark{1,2}, Vicky Kalogera\altaffilmark{1}, and Frederic A.~Rasio\altaffilmark{1}}
\affil{$^1$Center for Interdisciplinary Exploration \& Research in Astrophysics (CIERA)\\Physics \& Astronomy, Northwestern University, 
Evanston, IL 60202, USA\\sourav.chatterjee@northwestern.edu}
\affil{$^2$MIT-Kavli Institute for Astrophysics and Space Research\\77 Massachusetts Avenue, 37-664H, Cambridge, MA 02139, USA}

\begin{abstract}
Using numerical models for star clusters spanning a wide range in ages and metallicities 
($\metal$) we study the masses of binary black holes (BBHs) produced dynamically and 
merging in the local universe ($\redshift\lesssim0.2$). 
After taking into account cosmological constraints on star-formation rate 
and metallicity evolution, which realistically relate merger delay times obtained from 
models with merger redshifts, we show here for the first time that while old, metal-poor globular 
clusters can naturally produce merging BBHs with heavier components, as observed in 
GW150914, lower-mass 
BBHs like GW151226 are easily formed dynamically in younger, higher-metallicity clusters.
More specifically, we show that the mass of GW151226 is well within $1\sigma$ of the 
mass distribution obtained from our models for clusters with $\metal/\Zsun\gtrsim0.5$. 
Indeed dynamical formation of a system like GW151226 likely requires a cluster that is 
younger and has a higher metallicity than typical Galactic globular clusters. The LVT151012 
system, if real, could have been created in any cluster with $\metal/\Zsun\lesssim0.25$. 
On the other hand, GW150914 is more massive (beyond $1\sigma$) than typical BBHs 
from even the lowest-metallicity ($\metal/\Zsun=0.005$) clusters we consider, but is within 
$2\sigma$ of the intrinsic mass distribution from our cluster models with $\metal/\Zsun\lesssim0.05$; 
of course detection biases also push the observed distributions towards higher masses. 
\end{abstract}


\section{Introduction}
\label{S:intro}
Detection of gravitational waves (GWs) from merging black hole (BH) binaries 
has reignited widespread interest in understanding the astrophysical implications and the 
origins of BBHs \citep[][]{PhysRevLett.116.061102,PhysRevLett.116.241103,2041-8205-818-2-L22}. 
Current theoretical estimates indicate that detectable BBH merger events 
may be rather frequent, few--$500\,\mergerrate$, with large uncertainties depending on production channels and  
model assumptions (e.g., compare \citealt{2016PhRvD..93h4029R,2017MNRAS.464L..36A} with \citealt{2016MNRAS.460.3545D}). 
In the first observing run itself, the advanced LIGO observatories (aLIGO) have detected GW signals from  
two BBH mergers and a lower significance `trigger' event \citep{PhysRevLett.116.061102,PhysRevLett.116.241103,2016PhRvX...6d1015A}. 
These detections already show a large diversity in the masses of BBHs merging in the local 
universe: the chirp masses ($\mchirp$) at source for GW150914, LVT151012, and GW151226 are 
$28_{-2}^{+2}$, $15^{+1}_{-1}$, and $8.9^{+0.3}_{-0.3}$, respectively. 

Broadly speaking, two major channels have been proposed for BBH formation and subsequent merger. 
High-mass stellar binaries may evolve in isolation to create merging BBHs, for example, by going 
through a specific sequence of events involving low-kick supernovae (SNe) and common envelope 
(CE) evolution \citep[e.g.,][]{2012ApJ...759...52D,2013ApJ...779...72D,2015ApJ...806..263D,2014ApJ...789..120B,2016ApJ...819..108B,2015A&A...574A..58K,2016Natur.534..512B}, 
and via chemically homogeneous evolution of tidally distorted binaries 
\citep[e.g.,][]{2016MNRAS.458.2634M,2016A&A...588A..50M}. 

Alternatively, merging BBHs could be produced dynamically at the centers of 
dense star clusters \citep[e.g.,][]{2010MNRAS.402..371B,2014MNRAS.441.3703Z,2015PhRvL.115e1101R,2016PhRvD..93h4029R}. 
The process involved here is fundamentally different from BBH formation in isolation. 
A negligible fraction of BBHs formed in dense massive star clusters are primordial, and none with 
$\tdelay\geqslant1\,\gyr$ are composed of BHs formed from stars that were born in that 
binary \citep[e.g.,][]{2017ApJ...834...68C}. 
As massive stellar binaries evolve in dense star clusters, even if they were initially hard, mass loss 
from stellar winds and compact object formation can make these binaries soft. Consequently, stellar encounters 
and natal kicks during BH formation disrupt these primordial binaries. 
Later, single BHs dynamically acquire other BH companions via three-body binary formation 
and binary-mediated exchange interactions which preferentially insert the relatively more 
massive BHs into a binary ejecting a less massive non-BH 
member from it \citep[e.g.,][]{2003gmbp.book.....H,2012ApJ...754..152C}. Because of this, the BBH properties and 
their merger times, unlike those formed in isolation, do not depend on the assumptions of initial binarity or binary orbital 
properties. While the rate of mergers are affected by the assumptions for 
the IMF and natal kicks, the BBH masses and merger delay times ($\tdelay$) 
are insensitive to those as well \citep{2017ApJ...834...68C}.  

While several studies have modeled BBH formation in isolation for a wide range in metallicities taking 
into account the cosmological evolutions of the star formation rate ($\sfr$) 
and metallicity \citep[e.g.,][]{2016Natur.534..512B,2016MNRAS.461.3877D,2016MNRAS.460.3545D}, 
due to primarily the computational cost, numerical studies of dynamically formed BBHs 
have so far restricted themselves to either a narrow 
range in metallicities and ages typical of the Galactic globular clusters (GGCs)
\citep[e.g.,][]{2015PhRvL.115e1101R,2017ApJ...834...68C} or to low-mass 
(initial $N\sim5\times10^3$), young ($\sim100\,\myr$) star clusters \citep[e.g.,][]{2014MNRAS.441.3703Z}. 
These studies also assumed that all model clusters formed roughly at the same epoch independent of the 
metallicity 
to evaluate the redshifts of BBH mergers from $\tdelay$ found in the models.  
This of course is a simplification. Stars form with wide ranges in metallicities at any redshift \citep[e.g.,][]{2014ARA&A..52..415M}. 
Massive star clusters are also observed today with a large range in ages and metallicities, 
for example, in M51, M101, and the LMC 
\citep[e.g.,][]{2005A&A...431..905B,2006AJ....132..883B,2007A&A...469..925S}.
Even for the GGCs, the metallicity distribution has a long tail extending to $\Zsun$ \citep{1996AJ....112.1487H}.  

We relax past assumptions and consider BBH formation and merger in clusters spanning a wide range 
in metallicities and metallicity-dependent distributions for cluster-formation redshifts ($\zform$). Our goal is to 
investigate whether all hitherto detected GW sources could have been formed dynamically in star clusters. 
Furthermore, we study effects of star cluster metallicity and age on the detectable properties 
(mass and eccentricity) of BBH mergers. 
In \S\ref{S:numerical} we describe our numerical setup. In \S\ref{S:result} we show 
the key results. We conclude in \S\ref{S:conclude}. 

\section{Numerical Models}
\label{S:numerical}
We use our \henon-type Monte Carlo cluster dynamics code \cmc\ to model 
star clusters. \cmc\ includes all physical processes relevant 
to study BBH production, dynamical evolution, and mergers in star clusters 
\citep[e.g.,][]{2007ApJ...658.1047F,2010ApJ...719..915C,2013ApJS..204...15P,2016MNRAS.463.2109R}.
The initial structural properties are guided by those of the observed young massive clusters, 
thought to be similar in properties (except metallicity) to the progenitors of today's 
GCs \citep[e.g.,][]{2010ApJ...719..915C,2013MNRAS.429.2881C}. 
We use seven different metallicities spanning a large range: 
$\metal/\Zsun=0.005$, $0.025$, $0.05$, $0.25$, $0.5$, $0.75$, and $1$. 
Since we focus on studying the effects of star cluster metallicity on BBH mergers, 
we fix all other initial properties of our model clusters 
in the main set: all models initially have $N=8\times10^5$ single/binary stars. 
The initial positions and velocities are 
assigned following a King profile with $w_0=5$. The initial virial radius $r_v=2\,\pc$. 
The initial stellar masses (primary mass, $M_p$, in case of a binary) 
are drawn from the IMF given in \citet{2001MNRAS.322..231K} 
between $0.08$ and $150\,\msun$. 
The initial binary fraction is $f_b=10\%$. 
The secondary masses ($M_s$) are 
drawn from a uniform distribution between $0.08/M_p$ and $1$. The initial orbital 
periods for binaries are flat in logarithmic intervals, and the eccentricities ($e$) are thermal. The single and 
binary stellar evolution is modeled using the \sse\ and \bse\ software \citep{2000MNRAS.315..543H,2002MNRAS.329..897H} 
updated with state-of-the-art prescriptions for stellar winds \citep[e.g.,][]{2001A&A...369..574V} and 
fallback-dependent natal kick distribution for BHs \citep[e.g.,][]{2002ApJ...572..407B,2012ApJ...749...91F}. To 
improve statistics we repeat each model using different seeds.

We create two additional sets of models with $\metal/\Zsun=0.05$ by varying the initial 
$N$ and $r_v$ to study their effects. In one set we change the 
initial $r_v$ to $1\,\pc$. 
In the other, we vary the initial $N$ to $2\times10^5$ and 
$2\times10^6$. 
Relevant model properties are summarized in Table\ \ref{T:props}.

\section{Results}
\label{S:result}
\begin{figure}
\begin{center}
\plotone{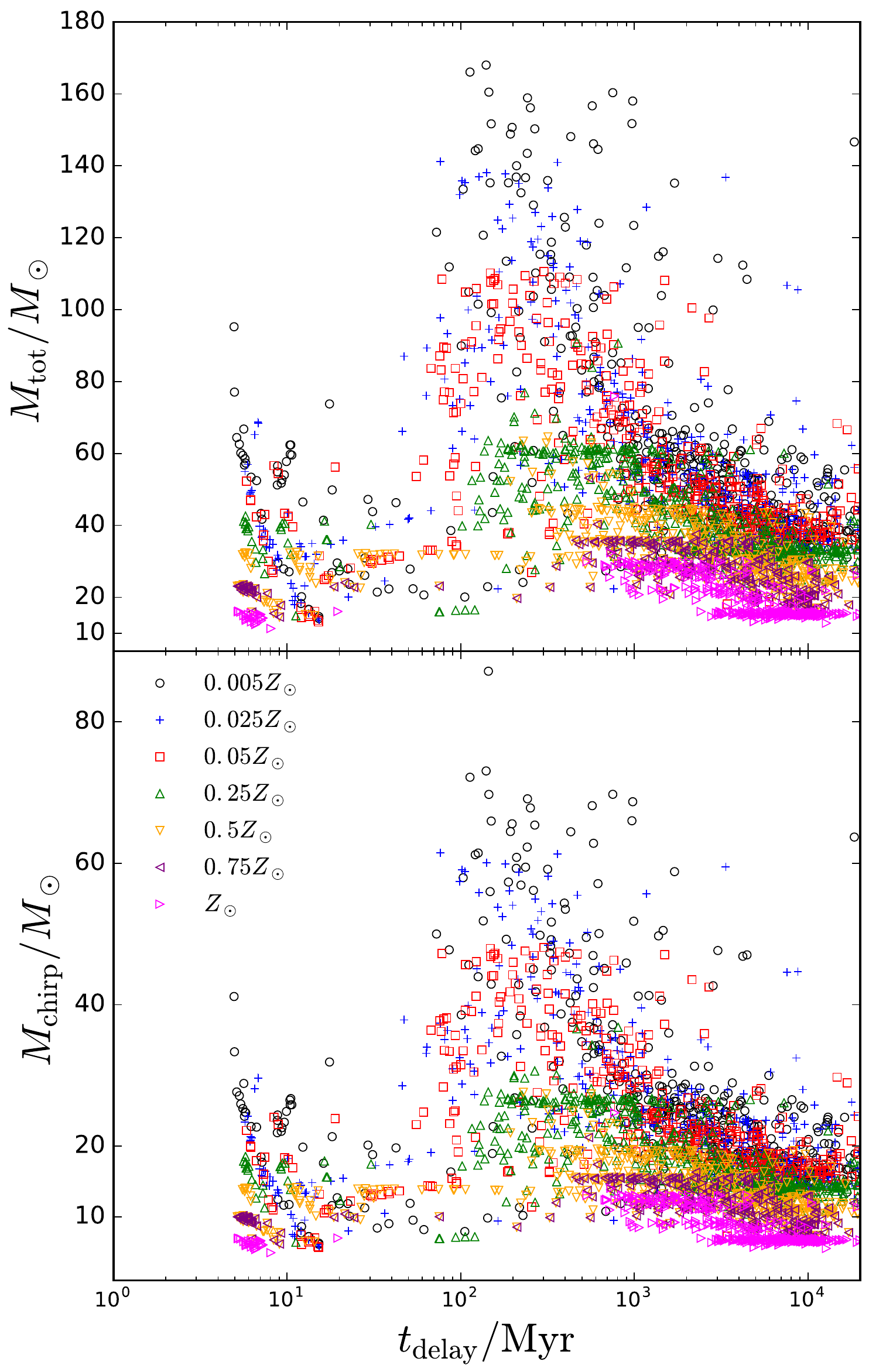}
\caption{
$\tdelay$ vs $\mtot$ (top) and $\mchirp$ (bottom) for BBH mergers from clusters modeled 
with different metallicities. 
Black (circle), blue (plus), red (square), green (triangle-up), 
orange (triangle-down), purple (triangle-left), and magenta (triangle-right) denote clusters modeled with 
$\metal/\Zsun=0.005$, $0.025$, $0.05$, 
$0.25$, $0.5$, $0.75$, and $1$, respectively. 
Merging BBHs from lower-metallicity clusters are more massive, a consequence of the $\metal$-dependence of 
the BH mass function at formation \citep[e.g.,][]{2010ApJ...715L.138B}. 
Heavier merging BBHs have shorter $\tdelay$ for any metallicity, a consequence of how 
BHs are dynamically processed inside clusters and the mass dependence of the inspiral time 
via GW radiation from a given initial separation \citep{1964PhRv..136.1224P}. 
Early ($\tdelay\lesssim100\,\myr$) mergers come from systems where both BH 
progenitors were in a primordial binary.
The apparent over-density of mergers at specific BBH masses for a given metallicity is due to 
spikes in the BH mass function at formation, expected from state-of-the-art 
progenitor-to-remnant mass relation \citep{2010ApJ...715L.138B}.    
}
\label{fig:tdelay}
\end{center}
\end{figure} 

Merging BBHs decrease in mass with increasing metallicity, a direct consequence of the metallicity-dependence 
of the BH mass spectrum at formation \citep[Fig.\ \ref{fig:tdelay};][]{2010ApJ...715L.138B}. 
Merging BBHs also decrease in mass 
as the merger delay time ($\tdelay$), defined by the time of BBH merger from $t=0$ for 
the cluster, increases. This is because clusters form and dynamically process higher-mass BBHs first, followed by 
less massive ones due to mass segregation \citep[][]{2013MNRAS.432.2779B,2015ApJ...800....9M}. 
Moreover, higher-mass BBHs merge 
faster due to GW radiation \citep{1964PhRv..136.1224P}.
 
Of all BBH mergers within a Hubble time, the fraction of in-cluster mergers 
varies between $\sim0.3$--$5\%$. 
The majority of 
all BBHs merge long after they are ejected from the cluster predominantly via 
dynamical scattering 
in the cluster's core. At most $\sim16\%$ of all mergers involve merging of 
BHs whose progenitors were initially members of the same binary. Even for these, 
either the BHs or their progenitors have had 
at least one strong encounter (typically many), such as binary-mediated scattering, and physical collisions before 
they merge \citep[similar conclusions in, e.g.,][]{2016ApJ...824L...8R}. 

\subsection{Merger time delay vs redshift}
\label{S:tdelay-to-z}
Connecting $\tdelay$ to merger redshift requires knowledge of the redshift of formation for the parent cluster. 
We closely follow the approach of \citet{2016Natur.534..512B} and 
adopt state-of-the-art cosmological constraints for $\sfr(\redshift)$ and $\metal(\redshift)$ 
to infer the $\zform$-distribution for clusters with a given metallicity. 
We adopt 
\begin{equation}
\sfr(\redshift) = 0.015 \frac{(1+\redshift)^{2.7}}{1+\{(a+\redshift)/2.9\}^{5.6}}\,\msun \Mpc^{-3} \yr^{-1} 
\label{eq:sfr}
\end{equation}
\citep{2014ARA&A..52..415M}. 
$\sfr(\redshift)$ peaks at $\redshift\simeq2$ (look-back time $\tlookback\simeq10\,\gyr$), 
and decreases 
by a factor of $5$ from its peak value by $\redshift=0.24$ ($\tlookback\simeq3\,\gyr$) 
and $\redshift=5.4$ ($\tlookback\simeq12\,\gyr$). 
The mean metallicity, $\Zmean$, is given by 
\begin{equation}
\log\Zmean(\redshift)=K+\log\left(\frac{y(1-R)}{\rho_b}\int_{\redshift}^{20}\frac{97.8\times10^{10}\sfr(\redshift')}{H_0 E(\redshift')(1+\redshift')}d\redshift'\right) 
\label{eq:meanZ}
\end{equation}
\citep[][their Eq.\ 2]{2016Natur.534..512B}. 
$R=0.27$ is the mass fraction of a generation of stars that remixes into 
the interstellar medium, $y=0.019$ is the net metal production, $\rho_b=2.77\times10^{11}\Omega_bh_0^2\,\dens$ 
is the baryon density, and $E(\redshift)=\sqrt{\Omega_M(1+\redshift)^3+\Omega_k(1+\redshift)^2+\Omega_\Lambda}$.
\footnote{We assume standard cosmological values: 
$\Omega_b=0.045$, $h_0=0.7$, $\Omega_\Lambda=0.7$, $\Omega_M=0.3$, $\Omega_k=0$, and $H_0=70\,\kms\Mpc^{-1}$.}  
We adopt normalization constant $K=1.30749$ to obtain $\Zmean=0.001$ and $0.02$ for $\tlookback\simeq12$ and $5\,\gyr$, 
respectively,  
guided by the typical ages and metallicities of the GGCs and 
the Sun. The exact adopted value of $K$ (within constraints) does not affect our results significantly. 

\begin{figure}
\begin{center}
\plotone{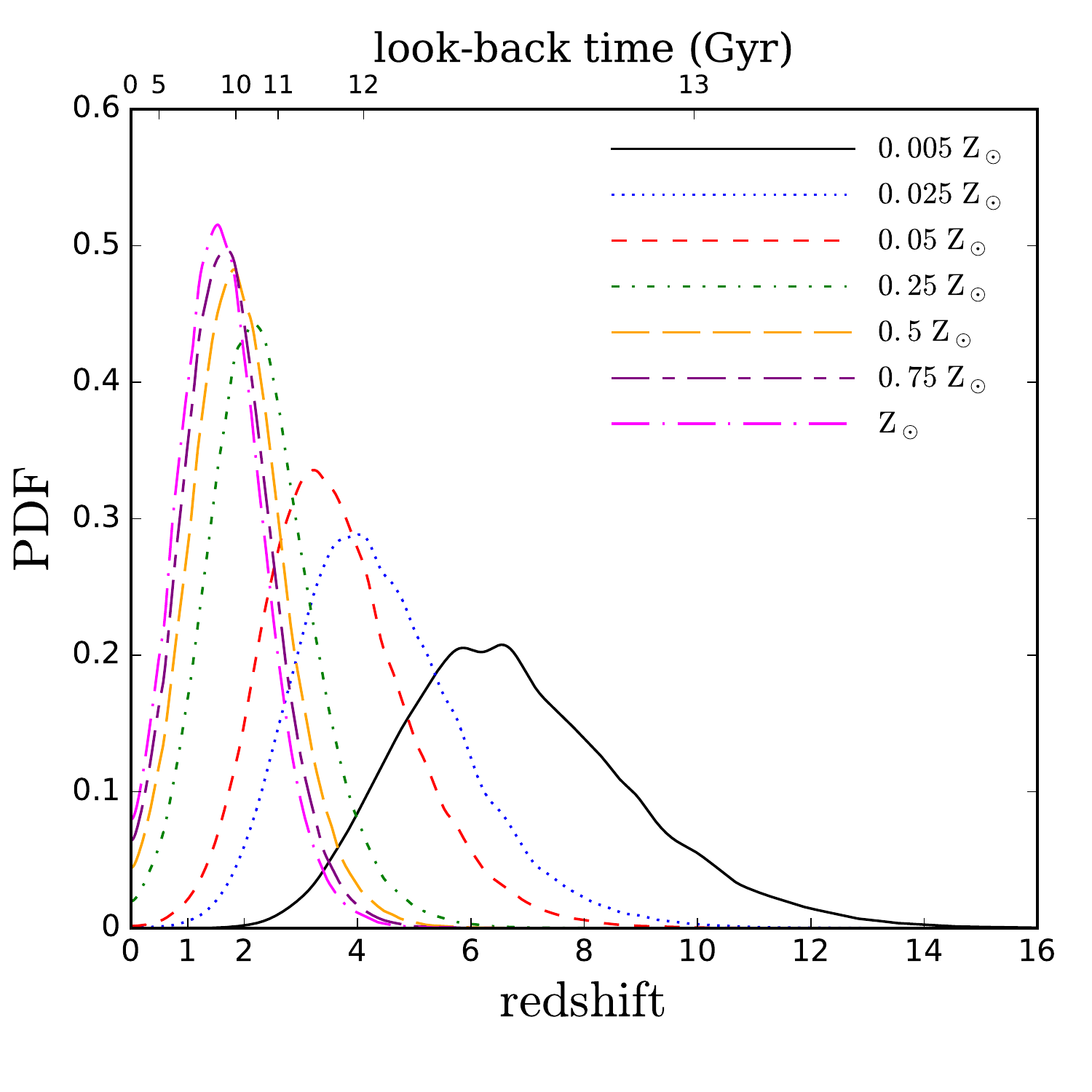}
\caption{
PDF for the redshift of formation ($\zform$) for star clusters of different metallicities (Eq.\ \ref{eq:fz}). 
Black (solid), blue (dotted), red (dashed), green (dash-dot), orange (long-dash), purple (long-dash-short-dash), 
and magenta (long-dash-dot) lines denote clusters modeled with metallicities 
$\metal/\Zsun=0.005$, $0.025$, $0.05$, $0.25$, $0.5$, $0.75$, and $1$, respectively. 
Look-back times corresponding to $\zform$ are also shown for reference. 
}
\label{fig:zdist}
\end{center}
\end{figure}

The probability distribution function (PDF) for $\zform$ of a given metallicity $\metal'$ is, 
\begin{equation}
f(\redshift)_{\metal'} = \int_{\metal=0.9\metal'}^{1.1\metal'} \int_{\redshift'=0}^{20} \sfr(\redshift') f'(\metal)_{\redshift'} d\redshift' d\metal, 
\label{eq:fz}
\end{equation}
where, $f'(\metal)_\redshift$ 
is assumed to be lognormal with $\sigma=0.5$ dex and mean$=\Zmean(\redshift)$, given by Eq.\ \ref{eq:meanZ} \citep{2016Natur.534..512B}. 
We evaluate $f(\redshift)_{\metal'}$ in the following way. 
We randomly generate $10^5$ redshift values between $0$ and $20$ weighted by $\sfr(\redshift)$ (Eq.\ \ref{eq:sfr}). 
For each draw of redshift we calculate 
$\Zmean$ and randomly generate $10^2$ metallicity values from $f'(\metal)_\redshift$.
Thus we generate a database of $10^7$ redshift-metallicity pairs. We then collect 
all redshift values corresponding to metallicities within $\metal'\pm0.1\metal'$. 
$f(\redshift)_{\metal'}$ is then obtained from these selected redshift values 
using a gaussian kernel density estimator 
(KDE) with bandwidth determined by Scott's method \citep{1992mde..book.....S}. 
We find that $f(\redshift)_\metal$ 
can be distributed across a large range in redshift, especially for clusters with low metallicities. 
Furthermore, due to the sharp peak of $\sfr(\redshift)$ 
at $\redshift\simeq2$, the modes of $f(\redshift)_\metal$, even for the highest metallicities we consider, 
are pushed towards $\redshift=2$ (Fig.\ \ref{fig:zdist}, Table\ \ref{T:props}). 

\subsection{Properties of BBH mergers within $0\leq z\leq1$}
\label{S:BBHprop}
We take $500$ random draws of 
$\zform$ from $f(\redshift)_{\metal'}$ (Eq.\ \ref{eq:fz}; Fig.\ \ref{fig:zdist}) 
for clusters of a given metallicity, $\metal'$.  
For each model of a particular metallicity and for each draw of $\zform$ we map the 
window of interest for BBH merger redshifts, $\Delta\redshift$ (e.g., $0\leq\redshift\leq1$), to the 
corresponding window in 
$\tdelay$, and collect all BBH mergers within $\Delta\redshift$. 
This essentially acts as a sliding 
window of selection of BBH mergers within $\Delta\redshift$ from a cluster  
based on the distribution of that cluster's formation times. 
We create a multi-dimensional (redshift, $M_p$, and $M_s$) PDF from the 
selected BBH mergers from all cluster models of a given set of initial properties. 
We then draw a sample of $10^5$ BBH mergers from this PDF 
to investigate the BBH merger properties for each metallicity.

\begin{figure*}
\begin{center}
\plotone{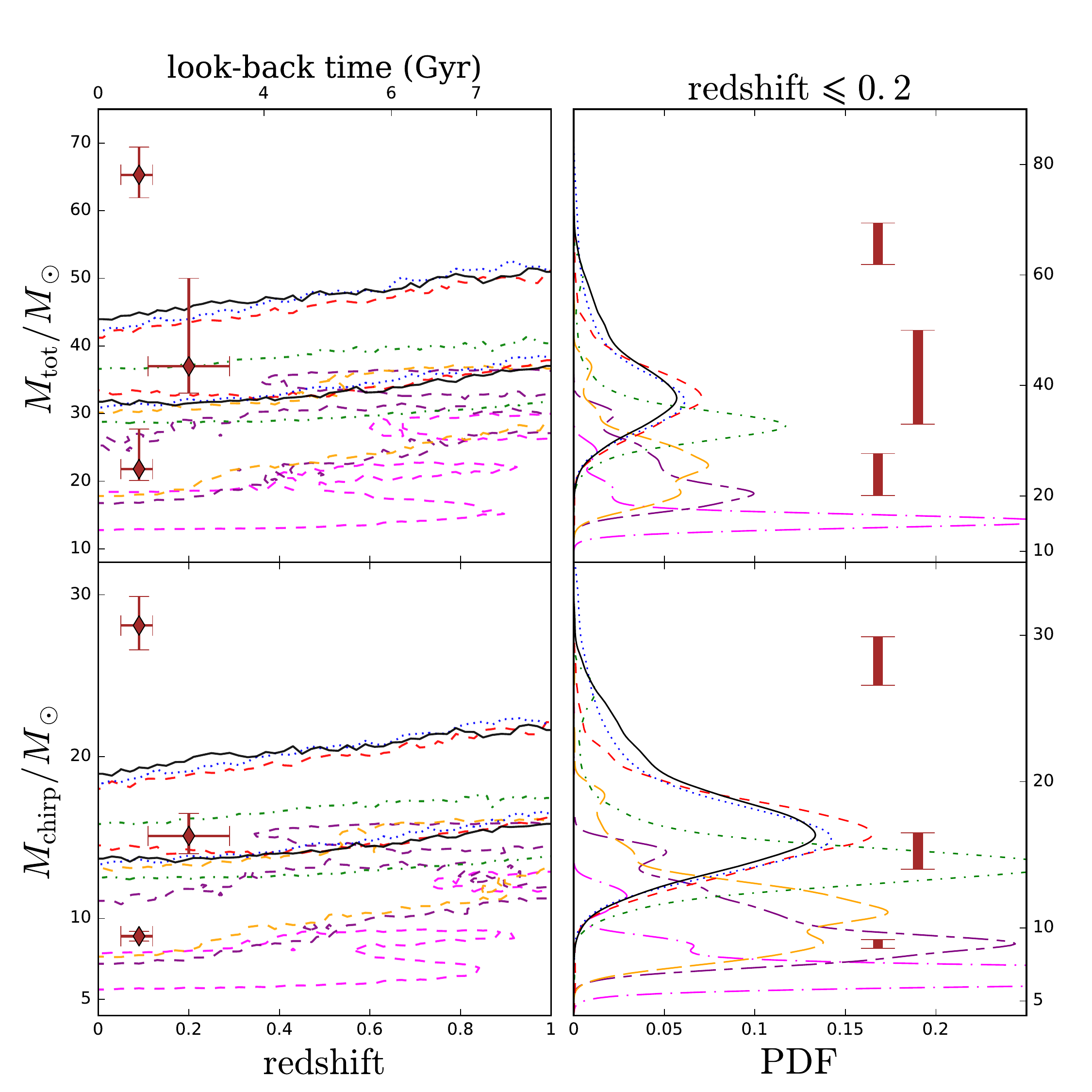}
\caption{
{\em Right}: mass distributions for BBHs merging in $\redshift\leq0.2$. 
{\em Left}: $1\sigma$ contours of the mass distributions of BBHs as a function of their 
merger redshift. Top and bottom panels show $\mtot$ and $\mchirp$. 
Line colors (and styles) have the same meaning as in Fig.\ \ref{fig:zdist}.
Brown diamonds and error-bars denote the source-frame properties and $90\%$ confidence 
intervals for the detected BBH mergers \citep{2016PhRvX...6d1015A}. 
}
\label{fig:zvsm}
\end{center}
\end{figure*} 
Fig.\ \ref{fig:zvsm} shows the distributions for $\mtot$ and $\mchirp$ 
for the merging BBHs from clusters with seven different metallicities. 
We find that the mass distributions are insensitive to the exact choice of $\sigma$ for $f'(Z)_z$ (\S\ref{S:tdelay-to-z}).
The left panels show $1\sigma$ contours for BBH mergers in the mass-redshift plane.  
The right panels show the mass distributions for BBHs merging in $\redshift\leq0.2$. 
For any metallicity, the $1\sigma$ contours 
encompass higher masses as redshift increases. 
This is because the lower the $\tdelay$ (equivalent to higher redshift), the higher the mass of 
merging BBHs (Fig.\ \ref{fig:tdelay}).
The mass distributions,  
even from clusters of vastly different metallicities, show significant overlap for any $\redshift\leq1$ 
(Fig.\ \ref{fig:zvsm}) primarily due to the wide ranges in merger masses for any 
$\tdelay$ and metallicity (Fig.\ \ref{fig:tdelay}). 

This makes uniquely inferring the metallicity of a particular merging BBH hard.  
Lower-metallicity clusters form higher-mass BBHs because BHs formed from lower-metallicity 
progenitors are heavier \citep[e.g.,][]{2010ApJ...715L.138B}. 
However, lower-metallicity clusters are older, hence,  
a particular observation window in redshift corresponds to higher $\tdelay$ and thus lower masses 
of merging BBHs (Fig.\ \ref{fig:tdelay}). 
For the same reasons, the mass distributions and their peaks for 
local BBH mergers 
do not change significantly for any 
$\metal/\Zsun\lesssim0.05$ (Fig.\ \ref{fig:zvsm}, Table\ \ref{T:props}). 

We compare the intrinsic mass distributions of BBH mergers from models 
with the source-frame $\mchirp$ of the detected GW sources. 
$\mchirp$ for LVT151012 and GW151226 are within $1\sigma$ of the $\mchirp$ 
distributions for BBHs merging in $\redshift\leq0.2$ from models with 
$\metal/\Zsun\leq0.25$ and $\metal/\Zsun\geq0.5$, respectively. 
Moreover, $\mchirp$ for LVT151012 and GW151226 line up perfectly with the peaks 
for $\metal/\Zsun=0.25$ and $0.75$, respectively (Fig.\ \ref{fig:zvsm}, Table\ \ref{T:props}).  

$\mchirp$ for GW150914 is within $2\sigma$ of the $\mchirp$ distributions 
for $\metal/\Zsun\leq0.05$, but is higher than $1\sigma$ of the $\mchirp$ distributions for any 
metallicities we consider. Since for $\metal/\Zsun\lesssim0.05$, 
$\mchirp$ distributions are not sensitive to the cluster metallicity (Fig.\ \ref{fig:zvsm}), 
GW150914 is likely more massive than intrinsically typical 
BBHs merging in $\redshift\leq0.2$. 
Accounting for detectability of BBH mergers by aLIGO, especially from low-metallicity clusters because 
the larger range of merger masses from them, 
would significantly reduce and enhance low- and high-mass regions of the PDF, respectively 
making GW150914 less rare among {\em detectable} BBH mergers from old, low-metallicity clusters 
\citep{2016ApJ...824L...8R}. 
Keeping this in mind, it is actually not surprising that the first 
ever detected GWs came from the merger of an intrinsically unusually massive BBH. 
BH dynamics in a star cluster increases 
BBH masses \citep[via repeated exchange encounters, e.g.,][]{2015PhRvL.115e1101R} and $\tdelay$ (Chatterjee et~al. 2016; in preparation) relative to BBHs formed in isolation. 
Thus, for a given metallicity and $\zform$, it is likely harder to create BBH mergers as massive as   
 GW150914 in isolation in the local universe. 

\subsection{Variation due to initial cluster properties}
\label{S:variation}
We now investigate how sensitive our results are on the initial cluster properties. 
The escape speed of the cluster, set by the cluster mass, sets the separation of the binaries at ejection, and thus 
$\tdelay$.  
The relaxation timescale ($\trelax$) controls the timescale for dynamical 
processing of BHs. Variations in initial $N$ and $r_v$ captures both of these effects. Other 
variations including initial binary fraction and binary orbital properties of high-mass stars, stellar 
IMF, distributions of natal kicks while can alter the merger rates, the mass distribution of 
BBHs merging in the local universe is insensitive to them \citep{2017ApJ...834...68C}. 

We find that the masses of local mergers from clusters of the same metallicity 
($\metal/\Zsun=0.05$) do not change significantly due to variations in initial $N$ and $r_v$. 
The modes of the distributions for both $\mtot$ and $\mchirp$ 
are well within $1\sigma$ of each other even when the initial $N$ is changed by an order of magnitude, 
and the initial $r_v$ is changed by a factor of $2$ (Table\ \ref{T:props}). 
Nevertheless, clusters with lower initial $r_v$ processes through the BHs quicker due to 
their shorter $\trelax$. Hence, the distribution is pushed towards slightly lower masses as $r_v$ is 
decreased (Table\ \ref{T:props}). 
On one hand, a higher-$N$ cluster (keeping all else fixed) ejects BBHs that are tighter, reducing 
$\tdelay$ \citep[e.g.,][]{2016PhRvD..93h4029R}. 
On the other hand, higher-$N$ increases $\trelax$, resulting in slower dynamical processing of 
BHs and slower decrease of $\mchirp$ with respect to $\tdelay$ 
\citep[Fig.\ \ref{fig:tdelay}; e.g.,][]{2015ApJ...800....9M}. 
These competing effects make the mass distributions for BBHs merging in $\redshift\lesssim0.2$ 
insensitive to the initial $N$ of the parent cluster. However, we caution that this trend should not be extrapolated 
to very low-$N$ clusters that dissolve before significant dynamical processing of their BHs, or to 
very high-$N$ clusters where $\trelax$ is longer than the cluster age.   

We have also followed the eccentricities of BBH orbits merging in $\redshift\leq1$ using the quadrupole 
approximated GW orbital evolution equations \citep{1964PhRv..136.1224P}. We find the $e$-distributions 
as the BBHs enter the 
aLIGO ($10\,\rm{Hz}$) and LISA ($10^{-4}\,\rm{Hz}$) frequency bands \citep[$\fgw$;][]{2003ApJ...598..419W}.
Similar to \citet{2016ApJ...830L..18B}, we find that at $\fgw=10\,\rm{Hz}$ BBHs have very low $e\sim10^{-7}$. 
Whereas, $e\sim0.01$--$0.4$ for the BBH orbits when $\fgw=10^{-4}\,\rm{Hz}$.\footnote{We neglect 
hierarchical triples which contribute at $\sim1\%$ level for clusters \citep{2016ApJ...816...65A}.}. 
We find no clear trends in the distributions of $\log e$ at these frequencies 
depending on metallicity, $N$ or $r_v$ (Table\ \ref{T:props}).    

\section{Conclusion}
\label{S:conclude}
We have studied the effects of the parent cluster's metallicity (and metallicity-dependent age) 
on the BBH masses merging in the local universe. 
Assuming cluster origin, we have found likely cluster properties of detected GW sources by 
comparing detected masses with mass distributions from models (\S\ref{S:BBHprop}, Fig.\ \ref{fig:zvsm}).
We find that $\mchirp$ of GW150914 is not within $1\sigma$ of the intrinsic $\mchirp$ distributions 
for BBHs merging in $\redshift\leq0.2$ for any metallicities we consider, but is within $2\sigma$ 
for mergers from clusters with $\metal/\Zsun\leq0.05$. Since below $\metal/\Zsun=0.05$ the 
$\mchirp$-distribution is insensitive to metallicity and dynamically created BBHs are generally heavier 
than those produced in isolation for any given metallicity, mergers of BBHs as massive as GW150914 
in $z\leq0.2$ are likely intrinsically rare. Of course, detection biases push the observed distributions 
towards higher masses. Since the lower the metallicity, the larger the range in merging BBH masses, 
detection biases would affect the mass distributions from lower-metallicity clusters more. 
Thus, the {\em detection} of mergers like GW150914 would be less rare \citep{2016ApJ...824L...8R}.   
$\mchirp$ of LVT151012 is near the peak of the distribution from clusters modeled with $\metal/\Zsun=0.25$, and is 
within $1\sigma$ of the distributions from all clusters modeled with $\metal/\Zsun\leq0.25$. 
$\mchirp$ of GW151226 is closest to the peak of $\mchirp$-distribution from clusters with $\metal/\Zsun=0.75$ and 
is within $1\sigma$ from clusters with $\metal/\Zsun\geq0.5$ 
Thus, assuming cluster origin, GW151226 likely formed in a higher-metallicity, younger cluster 
than typical GGCs. 

We find several additional notable trends. Less massive BBHs have longer $\tdelay$ for any metallicities 
(Fig.\ \ref{fig:tdelay}) since 
clusters dynamically form, process, and eject heavier BBHs earlier due to mass 
segregation \citep[e.g.,][]{2015ApJ...800....9M}. 
Lower-metallicity clusters typically have higher $\zform$ (Fig.\ \ref{fig:zdist}). 
Hence, BBHs from lower-metallicity clusters require longer $\tdelay$ to merge in $\redshift\lesssim0.2$. 
Lower metallicity leads to the formation of heavier BBHs \citep[e.g.,][]{2012ApJ...749...91F}, 
but longer $\tdelay$ decreases merging BBH masses (Fig.\ \ref{fig:tdelay}). Hence, 
while the expected trend is an increase of BBH masses merging in $\redshift\lesssim0.2$ as 
metallicity decreases, the mass distributions for local BBH mergers 
become insensitive to metallicity for $\metal/\Zsun\lesssim0.05$. 

Furthermore, we find that the masses of local BBH mergers 
are not very sensitive to the initial $N$ or $r_v$ even when $N$ is varied over an order of 
magnitude and $r_v$ by a factor of $2$ (Table\ \ref{T:props}). 
This indicates that the metallicity and metallicity-dependent age of the parent cluster are 
likely the most important properties to determine the peaks and distributions 
of BBH masses merging in the local universe. 

%
\acknowledgments{
This work was supported by NSF Grant AST-1312945, NSF Grant PHY-1307020, and NASA Grant NNX14AP92G. 
CR is grateful for the hospitality of the Kavli Institute for Theoretical Physics, supported by NSF Grant PHY11-25915, 
and is supported at MIT by a Pappalardo Fellowship in Physics. VK and FAR also acknowledge support from NSF 
Grant PHY-1066293 at the Aspen Center for Physics.
}


\clearpage
\onecolumngrid
\floattable
\rotate
\begin{deluxetable}{ccccc|cc|cc|cc|cc|cc}
\tabletypesize{\normalsize}
\tablecolumns{14}
\tablewidth{0pt}
\tablecaption{Star cluster models and BBH merger properties.}
\tablehead{
	  \colhead{$N$} &
	  \colhead{$M_i$} &
	  \colhead{$r_v$} & 
	  \colhead{$\metal/\Zsun$} &
	  \colhead{\#} &
	  \multicolumn{2}{c}{Cluster Formation Time} &
           \multicolumn{6}{c}{BBH Merger Properties} &
           \multicolumn{2}{c}{$N_{\rm{merge}}$} \\
           \cline{8-13}
           \colhead{($10^5$)} &
           \colhead{($10^5\ \msun$)} &
           \colhead{($\pc$)} &
           \colhead{} &
           \colhead{} &
           \colhead{$\zform$} & 
           \colhead{$\tlookback$} &
           \colhead{$\mtot$} &
           \colhead{$\mchirp$} &
           \colhead{$\mtot$} &
           \colhead{$\mchirp$} & 
            \colhead{$\log\ e_1$} &
	   \colhead{$\log\ e_{-4}$} &
            \multicolumn{2}{c}{} \\
           \colhead{} &
           \colhead{} &
           \colhead{} & 
           \colhead{} &
           \colhead{} &
           \colhead{} & 
           \colhead{($\gyr$)} &
	   \multicolumn{2}{c}{($\msun$)} &
           \multicolumn{2}{c}{($\msun$)} &
           \multicolumn{2}{c}{} &
           \colhead{} &
           \colhead{} \\
           \colhead{} &
           \colhead{} &
           \colhead{} & 
           \colhead{} & 
           \colhead{} &
           \colhead{} &
           \colhead{} &
           \multicolumn{2}{c}{$\redshift\leq0.2$} &
           \multicolumn{2}{c}{$\redshift\leq1$} &
           \multicolumn{2}{c}{$\redshift\leq1$} & 
           \colhead{$\redshift\leq0.2$} &
           \colhead{$\redshift\leq1$} \\
}
\startdata
\omit & \omit & \omit & $0.005$ & $4$ & $6.2_{-1.7}^{+1.9}$ & $12.6_{-0.4}^{+0.3}$ & $37.7_{-7.0}^{+8.4}$ & $16.4_{-3.1}^{+3.5}$ & $39.8_{-7.4}^{+9.4}$ & $17.3_{-3.3}^{+3.9}$ & $-7.3_{-0.8}^{+0.8}$ & $-2.0_{-0.8}^{0.8}$ & $4_{-1}^{+4}$ & $26_{-5}^{+2}$ \\
\omit & \omit & \omit & $0.025$ & $4$ & $4.2_{-1.4}^{+1.1}$ & $12.0_{-0.9}^{+0.4}$ & $37.2_{-6.8}^{+7.6}$ & $16.1_{-2.9}^{+3.3}$ & $39.2_{-6.5}^{+9.9}$ & $16.9_{-2.8}^{+4.2}$ & $-7.3_{-0.6}^{+1}$ & $-2.0_{-0.5}^{+1}$ & $4_{-1}^{+5}$ & $24_{-1}^{+7}$ \\
\omit & \omit & \omit & $0.05$ & $4$ & $3.4_{-1.1}^{+1.1}$ & $11.6_{-1.0}^{+0.5}$ & $37.6_{-5.9}^{+5.9}$ & $16.4_{-2.7}^{+2.5}$ & $39.6_{-6.9}^{+7.7}$ & $17.1_{-3.0}^{+3.4}$ & $-7.2_{-0.7}^{+0.8}$ & $-1.9_{-0.6}^{-0.8}$ & $4_{-2}^{+1}$ & $30_{-12}^{+2}$ \\
$8$ & $5$ & $2$ & $0.25$ & $4$ & $2.1_{-0.7}^{+1.0}$ & $10.3_{-1.5}^{+1.0}$ & $32.7_{-3.8}^{+3.8}$ & $14.2_{-1.7}^{+1.7}$ & $33.3_{-4.4}^{+6.4}$ & $14.5_{-1.9}^{+2.7}$ &  $-7.0_{-0.8}^{+1}$ & $-1.8_{-0.6}^{+1}$ & $6_{-3}^{+1}$ & $22_{-2}^{+3}$ \\
\omit & \omit & \omit & $0.5$ & $4$ & $1.8_{-0.7}^{+0.8}$ & $9.9_{-1.9}^{+1.1}$ & $25.6_{-6.9}^{+4.3}$ & $11.1_{-3.0}^{+1.8}$ & $27.8_{-5.8}^{+8.2}$ & $11.4_{-2.0}^{+4.2}$ & $-6.5_{-0.9}^{+0.8}$ & $-1.1_{-0.8}^{+0.7}$ & $3_{-1}^{+2}$ & $24_{-6}^{+1}$ \\
\omit & \omit & \omit & $0.75$ & $4$ & $1.6_{-0.7}^{+0.8}$ & $9.6_{-2.2}^{+1.2}$ & $20.5_{-3.5}^{+6.5}$ & $8.9_{-1.6}^{+2.6}$ & $28.5_{-7.5}^{+7.0}$ & $10.9_{-2.1}^{+4.4}$ & $-6.5_{-0.9}^{+0.8}$ & $-1.1_{-0.8}^{+0.7}$ & $5_{-3}^{+1}$ & $16_{-1}^{+5}$ \\
\omit & \omit & \omit & $1$ & $4$ & $1.6_{-0.8}^{+0.7}$ & $9.4_{-2.5}^{+1.1}$ & $15.4_{-1.7}^{+1.9}$ & $6.7_{-0.8}^{+0.8}$ & $15.5_{-2.2}^{+6.3}$ & $6.7_{-0.8}^{+2.5}$ & $-7.0_{-0.4}^{+1}$ & $-1.8_{-0.2}^{+1}$ & $8_{-4}^{+2}$ & $40_{-5}^{+1}$ \\
\hline
$8$ & $5$ & $1$ & $0.05$ & $4$ & $3.4_{-1.1}^{+1.1}$ & $11.6_{-1.0}^{+0.5}$ & $32.9_{-4.8}^{+4.8}$ & $14.1_{-2.2}^{+2.2}$ & $33.9_{-5.1}^{+6.1}$ & $14.5_{-2.1}^{+2.9}$ & $-6.4_{-1}^{+0.6}$ & $-1.0_{-1}^{+0.5}$ & $3_{-1}^{+3}$ & $22_{-1}^{+7}$ \\
$20$ & $12$ & $2$ & $0.05$ & $2$ & $3.4_{-1.1}^{+1.1}$ & $11.6_{-1.0}^{+0.5}$ & $41.8_{-4.6}^{+7.2}$ & $17.9_{-1.8}^{+3.4}$ & $48.3_{-8.6}^{+4.2}$ & $21.0_{-4.0}^{+1.8}$ & $-6.2_{-1}^{+0.7}$ & $-0.6_{-1}^{+0.5}$ & $16_{-6}^{+2}$ & $60_{-1}^{+9}$ \\
$2$ & $1$ & $2$ & $0.05$ & $18$ & $3.4_{-1.1}^{+1.1}$ & $11.6_{-1.0}^{+0.5}$ & $37_{-11}^{+30}$ & $16_{-4.8}^{+13}$ & $39_{-10}^{+26}$ & $17_{-5}^{+11}$ & $-6.7_{-1}^{+0.8}$ & $-1.4_{-1}^{+0.8}$ & $0_{-0}^{+4}$ & $3_{-1}^{+1}$ \\
%
%
\enddata
\tablecomments{$M_i$ is the initial cluster mass. 
\# denotes the number of models simulated with the same initial cluster properties. 
Cluster formation redshifts, $\zform$, and the equivalent look-back times, $\tlookback$, 
are shown for clusters of particular metallicities (\S\ref{S:tdelay-to-z}). 
We denote the eccentricities of BBH orbits (that merge in $\redshift\leq1$) when their 
GW frequency $f_{\rm{GW}}=10$ and $10^{-4}\,\rm{Hz}$ by $e_1$ and $e_{-4}$, respectively. 
$N_{\rm{merge}}$ denotes the number of BBH mergers. 
All numbers with error-bars denote the mode and $1\sigma$ range for the respective distributions. 
}
\label{T:props}
\end{deluxetable}

\end{document}